\documentclass[aps,prb,preprint,showpacs]{revtex4}
\usepackage{graphicx}
\usepackage{amssymb}
\begin{document}

\title{Interface superconductivity in La$_{1.48}$Nd$_{0.4}$Sr$_{0.12}$CuO$_{4}$/La$_{1.84}$Sr$_{0.16}$CuO$_{4}$ bilayers}

\author{P. K. Rout$^1$ and R. C. Budhani$^{1,2}$}
\email{rcb@iitk.ac.in}
\affiliation{$^1$Condensed Matter - Low Dimensional Systems
Laboratory, Department of Physics, Indian Institute of Technology,
Kanpur - 208016, India\\
$^2$National Physical Laboratory, New Delhi - 110012, India}

\date{\today}
\begin{abstract}
\baselineskip=1.5cm

We identify a distinct superconducting phase at the interface of a La$_{1.48}$Nd$_{0.4}$Sr$_{0.12}$CuO$_4$ (LNSCO)/La$_{1.84}$Sr$_{0.16}$CuO$_4$ (LSCO) epitaxial bilayer system using ac screening measurements. A model based on inter-diffusion of quasiparticles and condensate at the interface yields a thickness of $\sim$ 25 nm for the interfacial layer. Two-dimensional superconductivity of the interface layer appears to be  governed by Kosterlitz-Thouless-Berezinskii transition. A parallel magnetic field suppresses the superconducting transition temperature of this layer with a pair breaking parameter $\alpha$ varying as $H^2$.

\end{abstract}
\pacs{74.72.Gh, 74.78.-w, 74.25.nn, 74.25.Ha}
\maketitle

\section{Introduction}
\baselineskip=1.5cm

A range of novel physical phenomena like high mobility two-dimensional (2D) electron gas,\cite{Ohtomo} quantum Hall effect,\cite{Tsukazaki} magnetism,\cite{Brinkman} and interface superconductivity,\cite{Reyren, Gozar, Seguchi, Fogel} are observed at interfaces between complex oxides, which have been at the focus of intensive scientific research. The interfaces can be exploited to tailor unusual properties by modifying the electronic structure at the interface using controlling factors like interface roughness, strain due to lattice mismatch, structure reconstruction at the interface, charge depletion/accumulation due to the difference in chemical potentials, etc. Recently there have been many experiments showing the existence of superconductivity at the interfaces between two oxide insulators\cite{Reyren} due to charge accumulation on application of electric field and at metal-insulator interfaces\cite{Gozar} due to long-range electrostatic interactions. A dislocation-induced interface superconductivity has been reported in superlattices where one\cite{Seguchi} or both\cite{Fogel} components are semiconductors.

In this paper we present a study of bilayer thin film systems of optimally doped compound La$_{1.85}$Sr$_{0.16}$CuO$_4$ (LSCO) and static charge stripe ordered compound La$_{1.48}$Nd$_{0.4}$Sr$_{0.12}$CuO$_4$ (LNSCO) close to anomalous 1/8 doping level. This combination provides an interesting possibility of interaction between two contrasting phases in proximity to each other at the interface. We indeed observe a manifestation of this proximity in form of an additional superconducting (SC) interface layer in bilayers.

\section{Experiment}
The physical properties of doped La$_2$CuO$_4$ (LCO) epitaxial films are affected in a non-trivial manner by the single crystal substrate on which they are deposited. For example, La$_{1.9}$Sr$_{0.1}$CuO$_4$ films grown on (001) SrLaAlO$_{4}$ (SLAO) show a doubling of the superconducting transition temperature ($T_c$) from bulk value of 25 K to 49 K.\cite{Locquet} Similarly, thin films of La$_{1.6-x}$Nd$_{0.4}$Sr$_x$CuO$_4$ grown on (100) SrTiO$_3$ show insulating behavior as $T \rightarrow0$ while films on SLAO are superconducting with $T_c$ higher than the bulk value.\cite{Tsukada,Liu} We have chosen (001) SLAO as the substrate in order to have a bilayer with a distinct $T_c$ as well metallic behavior below 300 K. A multi-target pulsed laser deposition (PLD) technique based on KrF excimer laser ($\lambda$ = 248 nm) was used to realize a layer-by-layer growth of LSCO (100 nm) and LNSCO (100 nm) films as well as LNSCO ($d_n$ = 100 nm)/LSCO ($d_s$ = 50 nm) bilayers. The films and bilayers were grown at 800 $^{\circ}\mathrm{C}$ in 230 mTorr of oxygen pressure. The deposition chamber was filled with oxygen to atmospheric pressure after the growth and then the sample was cooled to room temperature with a 30 minutes holdup at 500 $^{\circ}\mathrm{C}$ to realize full oxygenation of the structure.

The crystallographic structure and interface quality of the films were characterized using X-ray scattering in $\theta$-2$\theta$, $\omega$, $\phi$ and grazing incidence X-ray reflectivity (GIXR) modes. The superconducting response of the films was measured through resistivity $\rho(T)$ as well as ac screening methods. For the former, the samples were patterned in four-probe geometry using photolithography. The strength of ac screening currents and their dissipative behavior in 3$\times$3 mm$^{2}$ films were measured with a two-coil mutual inductance method similar to the one described by Jeanneret {\it et al.},\cite{Jeanneret} in which an ac voltage is applied to the drive (primary) coil to create a magnetic field of amplitude $h_{ac} \approx$ 0.3 Oe and the voltage across the pick-up (secondary) coil is measured with a lock-in amplifier used in the differential mode. The measurements were performed at frequency $f$ = 100 kHz in the temperature range of 5-35 K. The ac response of the samples was also measured in the external dc magnetic field $H$ (0-3500 Oe) applied parallel to the plane of the sample.


\section{Result and discussion}

Figure 1(a) shows $\theta$-2$\theta$ X-ray diffraction pattern of a LNSCO (100 nm)/LSCO (50 nm) bilayer around the (008) and (0010) reflections of the SLAO substrate. We can see two overlapping Bragg peaks of which the lower angle component can be assigned to the reflections from (00$l$) oriented LSCO planes while the higher angle one is due to (00$l$) planes of the LNSCO. This assignment is based on the $c$-axis lattice parameters, $c_{LSCO}$ and $c_{LNSCO}$, which are 1.33 nm and 1.31 nm in the bulk form respectively. Rocking curves of the bilayers have a full width at half maximum of 0.2-0.4$^{\circ}$, indicating a high quality epitaxial growth. A typical rocking curve about (008) LSCO peak is shown in Fig. 1(b). A $\phi$-scan of the bilayer shows sharp peaks only at integral multiple of $\pi/2$ (Fig. 1(c)) indicating in-plane ordering of $ab$-plane. A fitting of the GIXR curve with a genetic algorithm\cite{Dane} yields a roughness of 3.3 nm at the LNSCO/LSCO interface.

Figure 2(a) shows the normalized in-plane resistivity $\rho _{norm} (T)$ = $\rho (T)$/$\rho$(275 K) for LSCO and LNSCO films. The LSCO film shows superconducting transition  at $T_{c}$ = 31.2 K with a broadening ($\Delta T_c$) of $\approx$ 3.4 K while the transition in LNSCO is characterized by $T_{c}$ = 12.1 K with $\Delta T_c \approx$ 3.1 K. Here the $T_c$ has been defined as the temperature where the resistance drops to 90 $\%$ of the extrapolated normal state value. In the case of LNSCO layer grown on SLAO (inset of Fig. 2(a)), the low temperature tetragonal phase seems to be suppressed as indicated by the absence of the resistivity jump at low temperature orthorhombic to low temperature tetragonal phase transition temperature seen at $T_d \approx$ 70 K in single crystals.\cite{Tranquada} The temperature dependence of resistivity of the LNSCO (100 nm)/LSCO (50 nm) bilayer is also shown in Fig. 2(a). The heterostructure displays a single SC transition at $T_{c}$ = 27.8 K, which is $\approx$ 3.4 K lower than the $T_c$ of the single layer LSCO film. We now consider the issue of this lowering of $T_c$ ($\approx$ 3.4 K) in bilayers and how the superconductivity in LNSCO, albeit with a much lower $T_c$, affects the response of the bilayer. The suppression of $T_c$ can be attributed to several factors including; (i) the tensile strain on LSCO due to LNSCO bottom layer. Quantitatively, the in-plane ($ab$) epitaxial strain $ \varepsilon _{ab} = (a_{LSCO}  - a_{LNSCO} )/a_{LSCO}$, dependence of $T_c$ for a tetragonal unit cell is given by\cite{Locquet} $T_c  = T_c (0) + 2(\delta T_c/\delta \varepsilon _{ab})\varepsilon _{ab}$. With $a_{LSCO}$ = 0.378 nm, $a_{LNSCO}$ = 0.380 nm, and $\delta T_c /\delta \varepsilon _{ab}$ = 325 K,\cite{Gugenberger} we have $\varepsilon _{ab}$ = 0.6 $\%$ and decrease in $T_c$ of $\approx$ 4 K, provided the strain is retained throughout the film thickness. This may not be true for a 50 nm thick film as the strain will tend to relax with increasing thickness. (ii) The $T_c$ in such films is also sensitive to oxygen content\cite{Bozovic} and may be a factor contributing to its suppression. (iii) The structural disorder in the film suppresses the areal superfluid density $n_{s}$ and hence the $T_c$. (iv) Lastly, we consider reduction in $T_c$ due to conventional proximity effect,\cite{de Gennes} given as $\Delta T_c/T_{c0} \approx 1.35 \xi^2_c(0)/d^2_s$, where $T_{c0}$ is the transition temperature of single layer SC film and $\xi_c$, the coherence length along $c$-axis ($\xi_c \ll d_s$). However a typical value of $\xi_c(0) \approx$ 0.7 nm\cite{Cyrot} yields negligibly small $\Delta T_c$ ($\approx$ 22 mK) compared to the observed suppression of $\approx$ 3.4 K.

A yet another effect of significant interest in the context of cuprate physics is the interplay between LSCO and LNSCO at the interface. However, any interfacial effect is difficult to detect with planar transport measurements as the superconductivity of LSCO electrically shorts the less conducting bottom layer of LNSCO. Here we show that the interface superconductivity is manifested prominently in the measurement of screening currents, provided the film thickness is smaller than its magnetic penetration depth. This condition is satisfied in our case as both $ab$-plane and $c$-axis penetration depths are much larger than the film thickness. The real and imaginary parts of the induced pick-up coil voltage $V$ due to ac screening current in LNSCO/LSCO bilayer are shown in Fig. 2(b). The real part of the response has two clear loss peaks at 9.5 K and 19.9 K, corresponding to SC transition of individual LNSCO and LSCO layers respectively. Corresponding to these peaks, two transitions reflected as a sudden change of slope of the imaginary part are also seen. It is clear from the figure that Im $V$ starts appearing after the realization of the zero-resistance state, which shows the screening measurements show the transition only when the whole sample becomes superconducting. A careful examination of the temperature dependence of Re $V$ reveals a small but distinct peak near $\approx$ 15.4 K located in between two prominent loss peaks. This feature provides a strong evidence for the existence of a third SC layer in the samples, which is presumably located at the interface of LSCO and LNSCO layers. It is instructive to see what are the other sources of this additional signal. For example, grain boundaries, ac field induced dissipation at weak links and flux motion in $ab$-plane can result in such additional peak.\cite{Nikolo} However, the single loss peak seen in our one-component monolayer films rules out the presence of such extraneous effects. Other possibility is related to inter-diffusion of Sr/Nd cations at the interface, which may result in a layer of its unique $T_c$. But the typical roughness of $\approx$ 3.3 nm at the LNSCO-LSCO interface as deduced from GIXR measurement, sets an upper limit to any intermixing and is quite small compared to the thickness over which the quasiparticle and Cooper pair correlations spread across the interface. Also a large volume of high resolution transmission electron microscopy studies on heterostructures of cuprates grown using PLD under similar conditions as used by us show atomically sharp interfaces.\cite{Triscone} Such studies suggest that the diffusion of cations under the condition of growth is limited to atomic distance only.

Since the appearance of the additional peak in the real part of pick-up coil voltage is the central result of this paper, it is important to comment on the likely reasons behind interface superconductivity seen in our bilayers. Appearance of this interface layer may be related to the role of stripes in LNSCO.\cite{Tranquada} In the normal state region of LNSCO films, we observe a crossover from metallic to insulator behavior in the resistivity curves. The insulating region is characterized by a logarithmic $T$ dependent resistivity given as $\rho \propto log(1/T)$, as shown in inset of Fig. 2(a). Similar transport behavior has been observed in single crystal LSCO systems below optimal doping,\cite{Komiya} in LNSCO single crystals,\cite{Noda} and in LNSCO thin films grown on SLAO.\cite{Liu} These observations suggest the possible existence of dynamically fluctuating stripes, if not static stripes. In the proximity of a robust condensate in LSCO, the confinement of charge carriers within dynamic charge stripes in LNSCO may be more relaxed due to the weakening of the pinning potential. Also at the same time, these stripes may induce an incipient dynamic stripe order in LSCO up to some distance from the interface. This intermediate stripe ordered state may be the reason behind the additional superconducting layer in our bilayers. Another possibility may be inferred from the theoretical studies of bilayer made up of underdoped and overdoped layers, where an enhancement of $T_c$ can be observed.\cite{Berg} The phase fluctuation dictates the $T_c$ in underdoped compounds while some sort of local superconducting pairing occurs without any phase coherence at higher temperatures. On the other hand, the pairing and phase order occur simultaneously, with a robust phase stiffness in overdoped compounds. If we make a bilayer of these two, then high phase stiffness of the overdoped layer phase locks the pairs in the underdoped layer via the interlayer tunneling. Although, in our case, LSCO is optimally doped; but it has a comparatively higher phase stiffness than LNSCO and thus it may introduce pairing correlation in LNSCO, which in turn creates a region with a new $T_c$ at the interface.

An estimate of the thickness of this interface layer can be made by considering two length scales relevant in this system; namely (1), the coherence length $\xi_n$ over which the transmission of Cooper pairs from a superconductor (S) to normal metal (N) occurs, as introduced in the context of proximity effect, and (2), a diffusion length $\Lambda$, which sets the scale for diffusion of quasiparticles from N in to S until they form Cooper pairs as explained by Blonder, Tinkham, and Klapwijk.\cite{Blonder} The conventional proximity effect\cite{de Gennes} gives $\xi_n \approx$ 0.5 nm for $c$-axis transport in our case. However it has been shown theoretically that the Josephson tunneling with barriers made up of a antiferromagnetic material\cite{Demler} or a phase-disordered superconductors in pseudogap state\cite{Marchand} is greatly enhanced compared to the conventional SNS tunneling. Giant proximity has also been observed experimentally in similar cuprate superconductor systems,\cite{Bozovic2} where the supercurrents can mediate within normal metal of thickness as large as $\approx$ 20 nm by resonant tunneling. In the light of these observations, the condensate of LSCO can indeed perturb LNSCO layer to greater depths from the interface. We now consider the diffusion of quasiparticles from LNSCO in to the top superconducting layer. The diffusion length is given as $\Lambda = \sqrt{D\tau_Q}$, where $D$ is the diffusion coefficient and the quasiparticle relaxation time, $\tau_Q \approx$ 6 ps.\cite{Kusar} The quasiparticle diffusion along $c$-direction is dominated by incoherent transport due to interlayer scattering.\cite{Turlakov} The diffusion coefficient is given as; $D = (c/2)^2/\tau_{hop}$, where $c$ is the $c$-axis lattice parameter of LSCO ($\approx$ 1.33 nm) and the interlayer hopping rate\cite{Tamasaku} is $1/\tau_{hop} \approx$ 180 cm$^{-1}$, from which we find $\Lambda \approx$ 3.8 nm.

In order to separate out the contribution of each layer to the net areal superconducting condensate density, we have deconvoluted the Im $V$ vs $T$ curve of the bilayer using measured single layer data, as shown in Fig. 3. The inset (b) of Fig. 3 shows the variation of $L^{-1}_{k}(T)$ vs $(1-(T/T_c)^2)$ for the three components of the nominal bilayer, with $T_c$ being obtained by extrapolating the linear high temperature portion of the $L^{-1}_{k}(T)$ vs $T$ curves (shown in inset (a)) to zero. Here inverse sheet kinetic inductance $L^{-1}_{k}(T)$ ($= d_s/\mu_0 \lambda_{ab}^{2}(T)$, where $\lambda _{ab} (T)$ is the magnetic penetration depth in $ab$-plane) is extracted from Im $V$ by a numerical inversion method\cite{Jeanneret} assuming Re $V \approx$ 0. The temperature dependence of $L^{-1}_{k}(T)$ clearly shows $\lambda _{ab} (T)$ follows a temperature dependence of the form $\lambda_{ab}^{- 2}(T) = \lambda_{ab}^{- 2}(0)[1-(T/T_c)^2]$ at low enough temperatures ($t < t^\ast$), which is a feature of $d$-wave superconducting order parameter modified by the effect of the disorders (which may be generated during film fabrication).\cite{Hirschfeld} It is noteworthy that $\lambda _{ab}(0)=$ 709 nm in case of LSCO layer is within the range of $\lambda _{ab}(0)$ (450-840 nm) found by various experiments on bulk as well as thin films.\cite{Paget} Moreover, the superconductor-to-normal phase transition is in accordance with quasi-2D thermal phase fluctuation model, where the measured $T_c$ is approximately the Kosterlitz-Thouless-Berezinskii (KTB) transition temperature for a single superconducting layer due to unbinding of vortex-antivortex pairs,\cite{BKT} and this temperature $T_{KTB}$ is related to kinetic inductance as:
\begin {eqnarray}
k_B T_{KTB}  = \frac{{\Phi _0^2 }}{{8\pi \mu {}_0}}\frac{d}{{\lambda_{ab}^2 (T_{KTB} )}}
\end {eqnarray}
where $\Phi _0$ is the flux quantum and $d$, the thickness of the layer. In Fig. 3(c), the intersection of the straight line, as predicted by Eq. (1), for CuO$_2$ planes coupled throughout the interfacial layer thickness with the measured $L^{-1}_{k}(T)$ yields $T_{KTB}$. This suggests that the superconductivity at the interface is essentially two dimensional in nature. Also we observe a continuous rapid drop in $L^{-1}_{k}(T)$ and thus in areal superfluid density $n_{s\square}$ of the interface layer. In theory, the superfluid density drops discontinuously to zero for a 2D superconductor. But weak interlayer coupling in cuprates softens this discontinuity.

The secondary coil voltages for three different in-plane dc magnetic fields are shown in Fig. 4(a$\&$b). The SC transition temperature of the interface layer, $T_c(H)$ decreases monotonically with increasing field while the imaginary part is suppressed. To identify the position of $T_c$ corresponding to loss peak maxima, we have fitted the Re $V$ vs $T$ curve with a three Gaussian function combination .\cite{Fit} In order to explain the magnetic field dependence of $T_c(H)$ for three layers shown in inset of Fig. 4(c), we have considered a pair-breaking model in gapless regime,\cite{de Gennes,Maki} given by the equation;
\begin {eqnarray}
ln \left[\frac{T_c(H)}{T_c(0)}\right] = -\frac{\pi \alpha}{4k_BT_c(H)}
\end {eqnarray}
in the limit $\alpha \ll k_BT_c(H)$, where $\alpha$ is the pair-breaking parameter. Since $\alpha$ has a quadratic dependence on field for thin films in parallel magnetic field,\cite{Maki} we have plotted $T_c(H) ln [T_c(H)/T_c(0)]$ as a function of $H^2$ in Fig. 4(c). An excellent fit seen here confirms the pair-breaking model for suppression of $T_c$. It is interesting to note that the loss peak due to interfacial layer almost disappears at fields in excess of 3000 Oe. The nature of the superconducting state of all three layers can be understood by the relationship between the magnetic field induced shift in $T_c(H)$ and the relative change of the inverse kinetic inductance $L^{-1}_{k}(0,H)/L^{-1}_{k}(0,0)$  in the limit $T\rightarrow$ 0, which shows a linear behavior within experimental accuracy (See Fig. 4(d)). This is in accordance with well known Uemura proportionality given as $T_c \propto n_{s\square}$.\cite{Uemura}

\section{Conclusion}
Bilayers of La$_{1.48}$Nd$_{0.4}$Sr$_{0.12}$CuO$_{4}$ and La$_{1.84}$Sr$_{0.16}$CuO$_{4}$ grown on (001) SLAO show a superconducting interface layer with critical temperature ($T_c \approx$ 15.4 K) intermediate between the $T_c$ of the La$_{1.84}$Sr$_{0.16}$CuO$_{4}$ and La$_{1.48}$Nd$_{0.4}$Sr$_{0.12}$CuO$_{4}$. Taking into account the inter-diffusion of quasiparticles and condensate at the interface, we estimated this layer to be $\sim$ 25 nm thick.  The superconducting order parameter of the interfacial layer appears to be of $d$-wave nature and this two-dimensional superconductivity follows Kosterlitz-Thouless-Berezinskii transition. The suppression of superconductivity due to the magnetic field ($H$) appears to be due to $H^2$ dependent pair-breaking. Here we have also established that the ac screening measurements can be used as a noble method to verify superconductivity at buried interfaces, which is not easily detectable by the conventional dc transport measurements.

\begin{acknowledgments}
The authors would like to thank Rajni Porwal for preparing the mutual inductance coils and P. C. Joshi for photolithography. P. K. Rout acknowledges financial support from Indian Institute of Technology (IIT) Kanpur and the Council for Scientific and Industrial Research (CSIR), Government of India.
\end{acknowledgments}

\clearpage

\begin{figure}[h]
\begin{center}
\includegraphics [width=9cm]{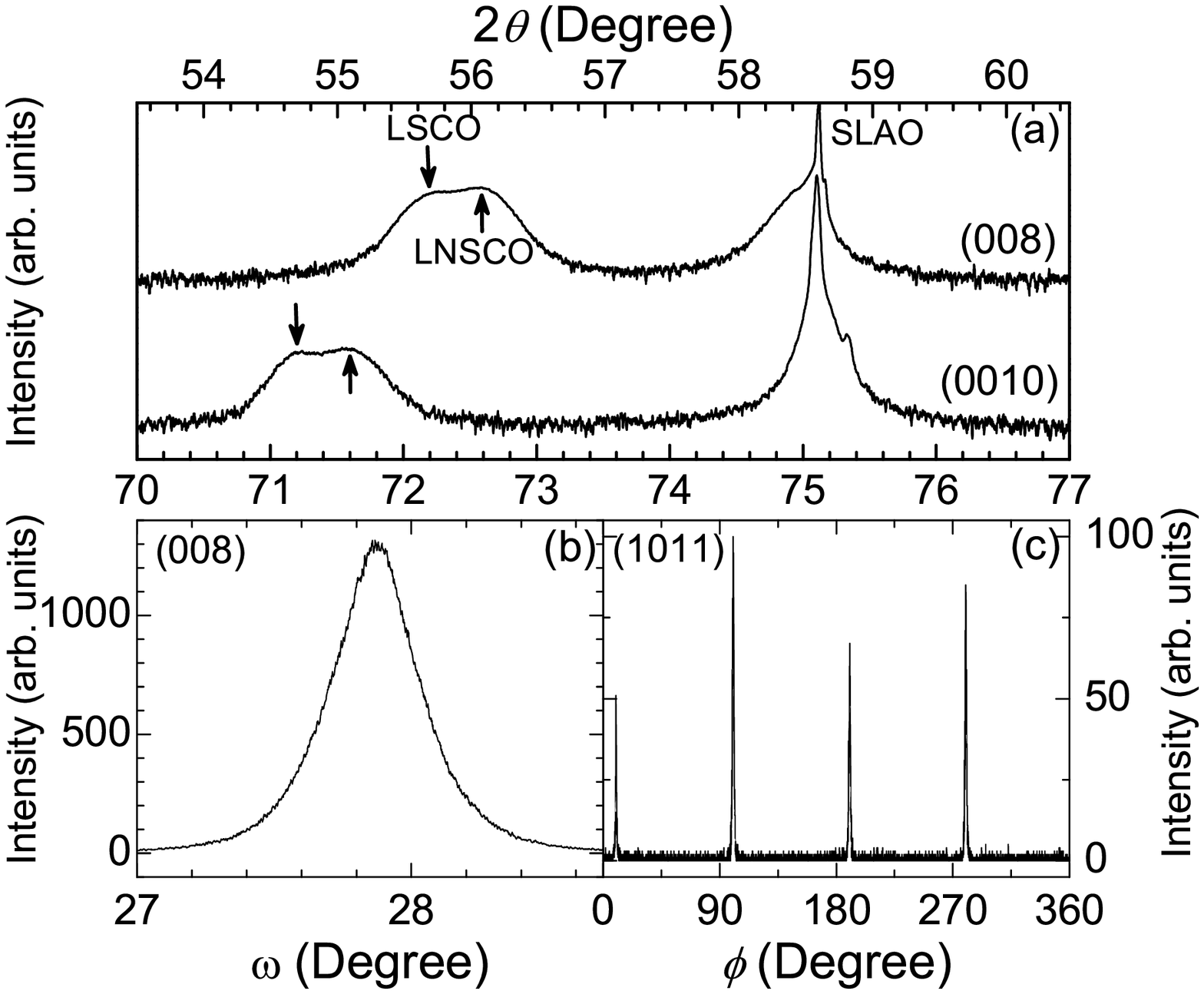}%
\end{center}
\caption{\label{fig1} Panel (a) shows $\theta$-2$\theta$ X-ray diffraction pattern of LNSCO (100 nm)/LSCO (50 nm) bilayer in the vicinity of the (008) and (0010) reflections of the SLAO (001) substrate. The position of LSCO peak is marked with a arrow pointing downwards whereas LNSCO peak is marked with up arrow. Panel (b) shows $\omega$-scan about (008) LSCO peak while $\phi$-scan of (1011) LSCO peak is shown in the panel (c).}
\end{figure}
\clearpage

\begin{figure}[h]
\begin{center}
\includegraphics [width=9cm]{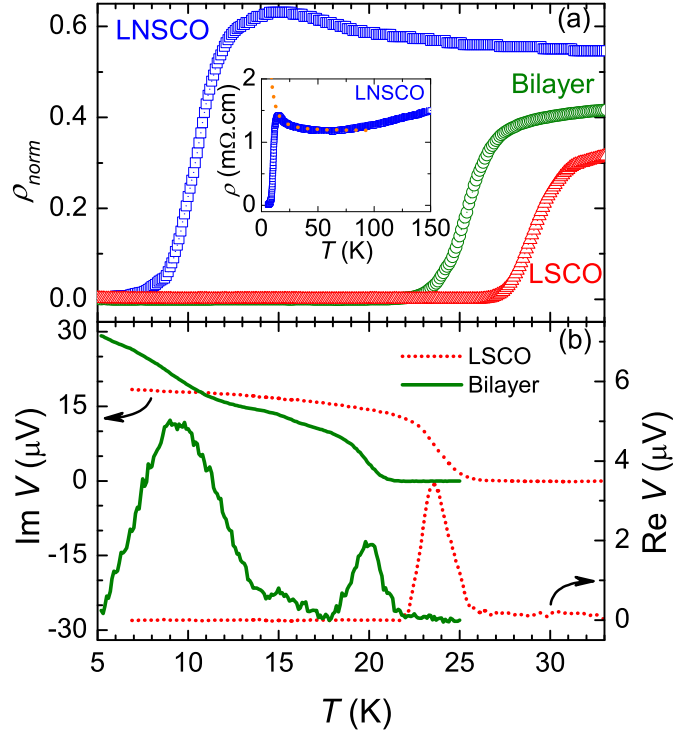}%
\end{center}
\caption{\label{fig2} (Color online)(a) Temperature dependence of normalized resistivity $\rho_{norm} (T)$ = $\rho (T)$/$\rho$(275 K) for LSCO and LNSCO films of thickness 100 nm each and of a LNSCO (100 nm)/LSCO (50 nm) bilayer. Inset shows $\rho(T)$ for LNSCO (100 nm) films on SLAO with the upturn of resistivity following the logarithmic $T$ dependence, as shown by dotted line. (b) Temperature dependence of the real and imaginary parts of the pick-up coil voltage for a 100 nm thick LSCO film showing one SC transition (Dotted line) and for a LNSCO (100 nm)/LSCO (50 nm) bilayer with three distinct transitions (Solid line).}
\end{figure}
\clearpage

\begin{figure}[h]
\begin{center}
\includegraphics [width=9cm]{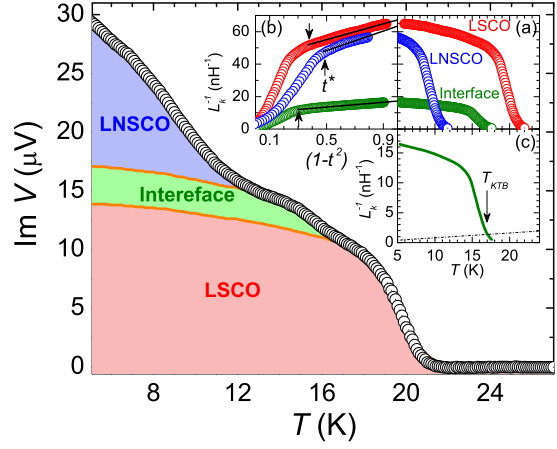}%
\end{center}
\caption{\label{fig3} (Color online)Main panel shows Im $V$ vs $T$ curve for LNSCO (100 nm)/LSCO (50 nm) bilayer with the contributions of LSCO, interface layer, and LNSCO shown by different shades. Inset (a) shows temperature dependence of inverse sheet kinetic inductance $L^{-1}_{k}(T)$ for three components of the bilayer separately. $L^{-1}_{k}(T)$ vs $(1-t^2)$ curves are shown in inset (b). Black straight lines show the $(1-t^2)$ dependence of $L^{-1}_{k}(T)$ at low temperatures ($t < t^\ast$). Here, $t = T/T_c$. Inset (c) shows KTB transition line (Dashed) for the interfacial layer along with corresponding $L^{-1}_{k}(T)$. }
\end{figure}
\clearpage

\begin{figure}[h]
\begin{center}
\includegraphics [width=9cm]{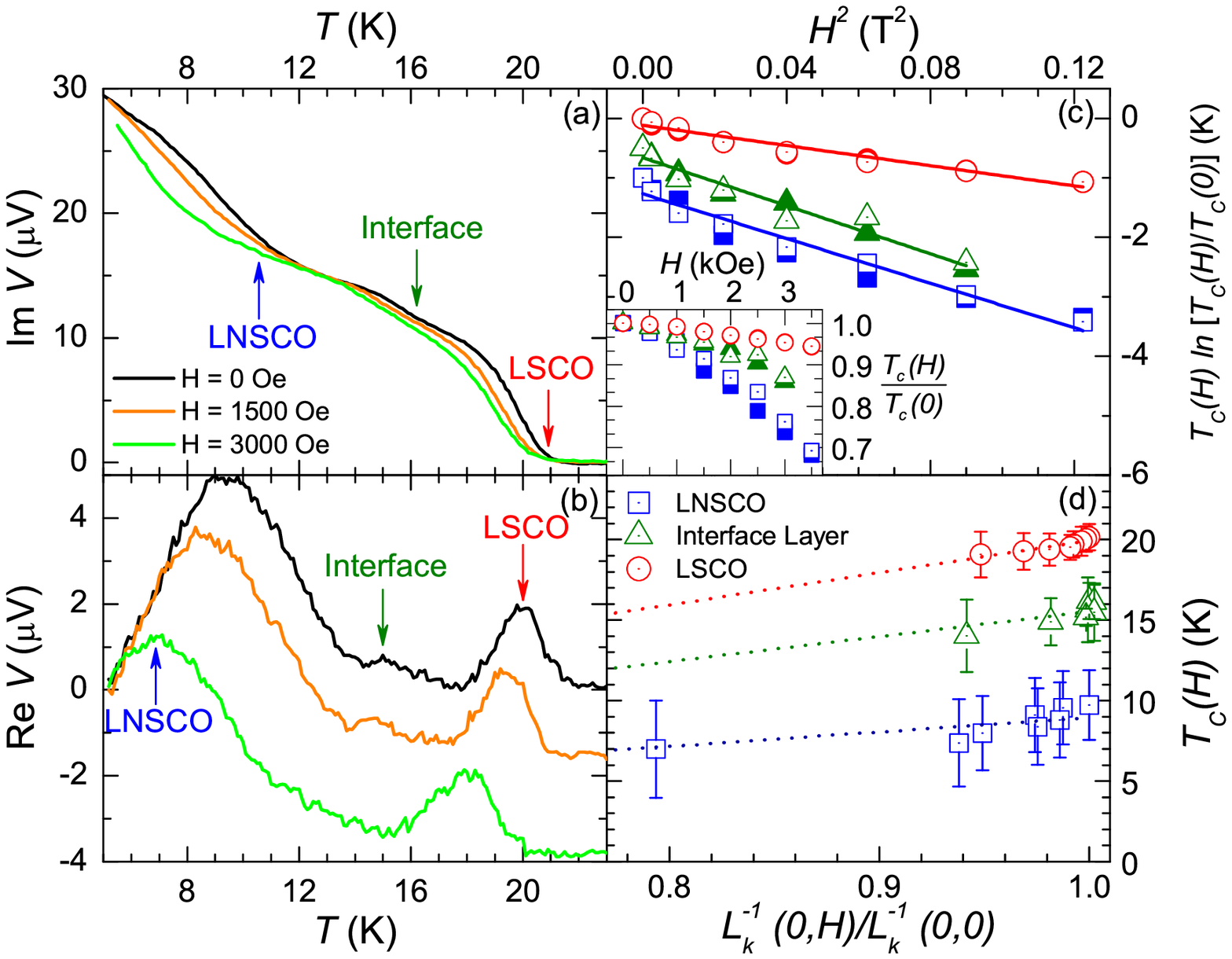}%
\end{center}
\caption{\label{fig4} (Color online)(a$\&$b) The pick-up coil voltage for LNSCO (100 nm)/LSCO (50 nm) bilayer is shown as a function of temperature for in-plane dc magnetic fields of 1500 Oe and 3000 Oe. Data for zero-field are also shown in the figure. (c) $T_c(H) ln [T_c(H)/T_c(0)]$ is plotted as a function of $H^{2}$ for three layers of LNSCO (100 nm)/LSCO (50 nm) bilayer with straight line fits given by Eq. (2). The curves are shifted down by 0.5 K from one another for clarity. Plot of $T_c(H)/T_c(0)$ vs $H$ is shown in the inset. Here, $T_c(H)$ values correspond to the temperature at which Re $V$ vs $T$ curves goes through a peak (Open symbols). We have also extracted $T_c(H)$ from the peak in $d$(Im $V$)/$dT$ vs $T$ curves (Filled symbols). (d) Plot of $T_c(H)$ vs $L^{-1}_{k}(0,H)/L^{-1}_{k}(0,0)$ with dotted straight line fit. Error bars shown correspond to transition width.}
\end{figure}


\begin{thebibliography}{100}
\bibitem{Fit}
Lorentzians were rejected because of the significant peak area associated with tails outside the transition region.
\end{thebibliography}
\end{document}